\pdfoutput=1
\documentclass[12pt]{article}

\usepackage{graphicx,psfrag,epsf,color}
\usepackage{amsmath,amssymb,amsfonts}
\usepackage{hyperref}
\usepackage{array}
\usepackage{bm} 
\usepackage{cite}
\usepackage{hyperref}
\usepackage{multirow}
\usepackage{rotating}
\usepackage{slashed}
\usepackage{textcomp}
\newcounter{MBQ}

\newcounter{KUQ}

\newcommand{\bff}[1]{\mbox{\boldmath ${#1}$}}
\newcommand{\grtsim}{\mbox{\raisebox{-3pt}{$\stackrel{>}{\sim}$}}}
\newcommand{\lessim}{\mbox{\raisebox{-3pt}{$\stackrel{<}{\sim}$}}}

\newcommand{\be}{\begin{equation}}
\newcommand{\ee}{\end{equation}}
\newcommand{\bea}{\begin{eqnarray}}
\newcommand{\eea}{\end{eqnarray}}
\newcommand{\bi}{\begin{itemize}}
\newcommand{\ei}{\end{itemize}}
\newcommand{\ben}{\begin{enumerate}}
\newcommand{\een}{\end{enumerate}}
\newcommand{\bt}{\begin{tabular}}
\newcommand{\et}{\end{tabular}}

\newcommand{\mchi}{m_\chi}

\newcommand{\vev}{\mathfrak{v}}

\begin{document}
\allowdisplaybreaks

\begin{titlepage}

\begin{flushright}
{\small
TUM-HEP-1419/22\\
arXiv:2209.14343 [hep-ph]\\[0.0cm]
January 03, 2023
}
\end{flushright}

\vskip1cm
\begin{center}
{\Large \bf 
Sommerfeld enhancement of resonant dark\\[0.0cm] 
 matter annihilation \\[0.2cm]}
\end{center}

\vspace{0.45cm}
\begin{center}
{\sc M.~Beneke$^{a}$, S.~Lederer$^{a}$,} and {\sc K.~Urban$^{a}$}\\[6mm]
{\it ${}^a$Physik Department T31,\\
James-Franck-Stra\ss e~1, 
Technische Universit\"at M\"unchen,\\
D--85748 Garching, Germany}
\end{center}

\vspace{0.55cm}
\begin{abstract}
\vskip0.2cm\noindent
The dark matter annihilation cross section can be amplified by 
orders of magnitude if the annihilation occurs into a narrow 
resonance, or if the dark-matter particles experience a 
long-range force before annihilation (Sommerfeld effect). We 
show that when both enhancements are present they factorize 
completely, that is, all long-distance non-factorizable effects 
cancel at leading order in the small-velocity and narrow-width 
expansion. We then investigate the viability of 
``super-resonant'' annihilation from the coaction of both 
mechanisms in Standard Model Higgs portal 
and simplified MSSM-inspired dark-matter scenarios.  
\end{abstract}
\end{titlepage}

\section{Introduction}
\label{sec:introduction}

In many models, the dark matter (DM) annihilation cross section, which constitutes the fundamental quantity determining the relic abundance and cosmic ray signals of DM, is enhanced beyond the naive expectation derived from the mass and couplings of the DM particle. The two most common enhancement mechanisms are non-relativistic scattering before pair annihilation, the Sommerfeld effect \cite{Sommerfeld,Hisano:2004ds,ArkaniHamed:2008qn}, and annihilation through a narrow resonance \cite{Ibe:2008ye,Guo:2009aj}. The Sommerfeld effect is generic for DM with SU(2) electroweak charge and mass in the TeV region and exhibits resonant behaviour for specific mass values. Resonant annihilation requires the resonance mass to be close to twice the DM mass, and can significantly boost the annihilation of weakly coupled DM. Interesting scenarios of this type can be found in the parameter space of the minimal supersymmetric standard model (MSSM) and Higgs portal models.

\begin{figure}[b]
\centering
\includegraphics[width=0.42\textwidth]{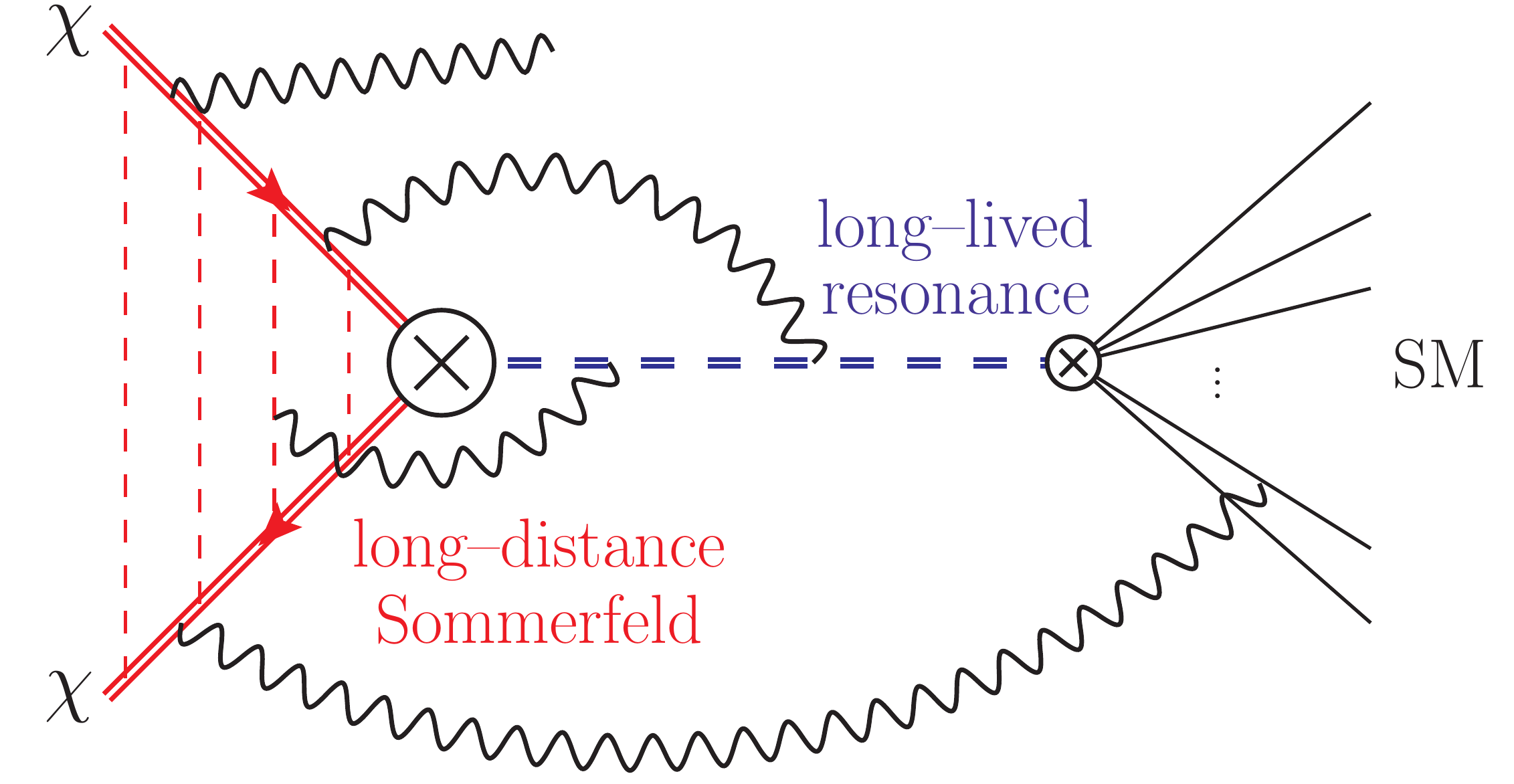}
\caption{\small Schematic representation of the process in question. In red, the exchange of a light mediator that 
produces the long-distance Sommerfeld effect, in blue 
the long-lived intermediate resonance. The question arises 
whether there are long-distance non-factorizable corrections connecting the two (black curly lines, including real 
emission into the final state).}
\label{fig:examplediag}
\end{figure}

The present work is motivated by two theoretical questions that appear when both effects, Sommerfeld and resonant enhancement, occur simultaneously. In this case the annihilation process is not only no longer a short-distance process, but it contains two long-distance phenomena possibly intertwined: a long-range force before annihilation, and a long-lived intermediate state thereafter, as sketched in Figure~\ref{fig:examplediag}. 
Can one factorize the two and achieve a simple description as is the case when only one of the enhancements is present?\footnote{The factorization of the Sommerfeld from another 
long-distance effect holds for the electroweak Sudakov 
resummation, see for example \cite{Beneke:2019vhz}.} Furthermore, the velocity-dependent Sommerfeld factor is then to be folded with an annihilation cross section into the resonance, which varies rapidly with the three-momentum of the DM particle. Does this lead to a breakdown of the partial-wave expansion in the calculation of the Sommerfeld effect?

The answer to both questions turns out to be simple, as will be seen, but to our knowledge has not been discussed previously in the literature. We exemplify the theoretical discussion with the Standard Model (SM) Higgs portal model augmented by a light mediator in the dark sector, and a MSSM-like model with TeV scale DM annihilating through a heavy Higgs boson resonance. 


\section{Factorization of the Sommerfeld and resonant enhancement}

The salient features of simultaneous Sommerfeld- and 
resonance-enhanced annihilation can be exposed by a simple model of a 
spin-1/2 DM field $\chi$ with gauge interactions 
in a representation of some gauge group $G$. The DM  
field couples to a resonance $H$, which may (but does not 
have to) transform under the same gauge symmetry, through 
the interaction 
\begin{equation}
\lambda \,y_{abc}\bar{\chi}_a\chi_b H_c + \mbox{h.c.}\,,
\label{eq:interaction}
\end{equation} 
where $a,b,c$ denote gauge-group indices, $\lambda$ is the 
coupling strength, and 
$y_{abc}$ are Clebsch-Gordan coefficients such that the 
product is gauge-invariant. We assume that the gauge 
symmetry is spontaneously broken such that the gauge 
boson acquires a small mass~$m_X$.

We are interested in a situation, where the DM  
particles are non-relativistic and twice the DM mass,  
$2m_\chi$, is close to the resonance mass $m_h$, and  
define 
\begin{align}
\delta M = m_h-2 m_\chi \ll m_h\,,
\end{align}
where $|\delta M|\ll m_h$. 
In the centre-of-mass (cms) frame of the annihilating 
$\chi\chi$ pair, the kinetic energies of the DM particles 
and resonance are small. The problem involves the 
high-mass, ``hard'' scales $m_\chi$ and $m_h$, as well as the 
low-energy scales $\delta M$, the width of the resonance 
$\Gamma_h$, and the kinetic energies. 

One can integrate out systematically the hard scales 
to construct an effective description of the annihilation 
process in terms of an effective Lagrangian. The 
resonance dynamics is described by the scalar field 
version of a static field (similar to heavy quark 
effective theory), $h_w$, 
generalized to an unstable particle 
\cite{Beneke:2003xh,Beneke:2004km}, given by
\begin{align}
\mathcal{L}_{\rm HSET} = h_w^\dagger 
\left( i D_0  - \delta M+ \frac{i \Gamma_h}{2} \right) h_w\,.
\label{eq:HSET}
\end{align}
$D_0$ is the time-component of the $G$-covariant 
derivative and 
the quantity $\delta M$ appears, since we measure the 
non-relativistic energy $E$ in the rest frame of the $\chi\chi$ pair, that is, from $2\mchi$ rather than from 
$m_h$. In this frame $w^\mu=(1,\mathbf{0})$. The Lagrangian 
receives small corrections suppressed by powers of 
$E/m_h, \Gamma_h/m_h$, which will be consistently 
neglected. 

The non-relativistic dynamics of the DM field $\chi$ is 
described by potential non-relativistic dark matter (PNRDM) 
effective theory \cite{Hisano:2004ds,Beneke:2014gja}
with the Lagrangian
\begin{align}
\mathcal{L}_{\rm PNRDM} = \chi_w^\dagger 
\left(i D_0 + \frac{\bff{\partial}^2}{2 m_\chi} \right) \chi_w 
- \int d^3 \mathbf{r} \, V(r) \left[\chi_w^\dagger \chi_w\right](t, \mathbf{x}) \left[\chi_w^\dagger \chi_w\right](t, \mathbf{x} + \mathbf{r})\,.
\label{eq:PNRDM}
\end{align}
Gauge-boson exchange generates the static Yukawa 
potential $V(r)$
\begin{align}
V(r) = - \frac{\alpha_X}{r} e^{-m_X r} \,,
\end{align}
which leads to Sommerfeld-enhanced DM annihilation. 
We denote the non-relativistic DM field by the symbol 
$\chi_w$ to distinguish it from $\chi$ above. We also assumed 
that the $\chi_a\chi_b$ scattering amplitude has been 
decomposed in irreducible group representations 
and that the corresponding Casimir has been absorbed 
into the coupling $\alpha_X$. In the following, we consider 
a single representation with an attractive potential 
and therefore drop the gauge index on the 
non-relativistic $\chi_w$ field.

The Sommerfeld effect is a long-distance effect related 
to the range $1/m_X$ of the potential. For specific 
values of $m_\chi, m_X$ and the gauge coupling 
a zero-energy DM bound-state appears in the spectrum.  
At these values the annihilation cross section is resonantly enhanced 
and grows as $1/v^2$ with the cms velocity $v$ of 
the annihilating particles. Similarly, the annihilation 
through a resonance is an enhanced long-distance effect 
relative to the point-like annihilation cross section, 
related to the life-time $1/\Gamma_h$ of the resonance. 
For small width $\Gamma_h$ and $\delta M=0$, the resonance 
enhancement grows as $1/v^4$. In the following, we 
first consider the factorization of both effects and 
then discuss their interplay.   

The inclusive DM annihilation cross section 
can conveniently be obtained from the discontinuity of the 
forward-scattering amplitude $\mathcal{T}$ through the 
optical theorem via $\sigma v_{\rm rel} = \frac{1}{m_\chi \sqrt{s}} 
\,\mbox{Im}\,\mathcal{T}$, with $\sqrt{s}=2m_\chi+E$ 
the cms energy of the process and $v_{\rm rel}=2 v$.
Unstable-particle effective theory \cite{Beneke:2003xh,Beneke:2004km} 
provides the systematic 
framework to describe resonant processes in an 
expansion in $\Gamma_h/m_h$. The forward-scattering amplitude 
of two non-relativistic DM particles can be expressed as 
\begin{align}
i \mathcal{T} = \sum_{m,n}\int d^4 x \,
\langle \chi \bar \chi|T\{ i J_m^\dagger(x) iJ_n(0)\} | \chi \bar \chi \rangle + \sum_k \left\langle \chi \bar \chi \middle|i T_k(0) \middle| \chi \bar \chi \right\rangle\,,
\label{eq:matrix_elem}
\end{align}
where the first term involves production operators 
$J_n$ for the long-lived resonance, whilst the second 
captures non-resonant local interactions through 
operators~$T_k$. This second term is suppressed relative 
to the first by one power of $\Gamma_h/m_h$, and can  
therefore be dropped for the following leading-power 
analysis.\footnote{The local term naturally factorizes by the 
standard non-relativistic analysis in the absence of a 
resonance, see, e.g., the discussion for the 
neutralino in \cite{Beneke:2012tg}.} For the simple model 
above, the production operator is 
\begin{align}
J(x) = \frac{C}{\sqrt{2 m_h}} y_{abc} \,(\chi_{wa}^\dagger\chi_{wb} h_{wc})(x)\,,
\end{align}
where $C$ denotes the hard matching coefficient 
and $C=\lambda$ at tree level. 

We now show that at leading power in the expansion in 
$\Gamma_h/m_h$ and the non-relativistic velocity, the 
long-distance dynamics before annihilation (Sommerfeld 
effect) is completely factorized from the long-lived 
resonance, that is, there are no soft gauge boson exchanges 
connecting the initial state to the resonance or the 
SM final state. To this end, we redefine the effective 
fields as 
\begin{equation}
h_w(x) \to Y_w(x_0) h_w(x), \qquad 
\chi_w(x)\to Y_w(x_0) \chi_w(x), 
\end{equation}
where 
\begin{align}
Y_w(x) = \mbox{P} \exp\left(i g_X \int_{-\infty}^0 \!ds\, 
w\cdot A^d_{s}(x + s w)\, T^d\right) 
\end{align}
is a time-like Wilson line, and the space-time point 
$x_0$ is $(t,\mathbf{0})$. The generator $T^d$ is taken 
in the representation of the field to which the Wilson 
line is multiplied. The Wilson line satisfies the 
key property 
\begin{equation}
Y_w^\dagger \,iw\cdot D Y_w = i w\cdot\partial\,.
\label{eq:Wilsonidentity}
\end{equation}
Inserting the above field redefinitions into 
the Lagrangians \eqref{eq:HSET}, \eqref{eq:PNRDM} 
removes the gauge field from the covariant derivative due to
the identity \eqref{eq:Wilsonidentity} and also leaves the 
Yukawa potential interaction 
invariant \cite{Beneke:2010da}. Thus the Lagrangians 
take the same form as before except for $D_0\to \partial_0$. 
The field redefinition puts the gauge interactions 
into the production vertex,  
\begin{equation}
y_{abc} \,(\chi_{wa}^\dagger\chi_{wb} h_{wc})(x)
\to y_{abc} \left[Y^{\dagger}_{w,aa^\prime} Y_{w, bb^\prime} 
Y_{w,cc^\prime}\right]\!(x_0) \,
(\chi_{wa^\prime}^\dagger\chi_{wb^\prime} h_{wc^\prime})(x)
\,.
\end{equation}
However, since $Y_w(x_0)$ is simply a gauge transformation 
with gauge parameter  $\alpha^d(x_0) = g_X\int_{-\infty}^0 \!ds 
\, w \cdot A^d_s(x_0 + s w)$, it follows from the 
gauge invariance of the production vertex 
that the right-hand side 
is in fact equal to the left. This demonstrates that 
gauge interactions are completely absent in the low-energy 
effective theory and hence all non-factorizable effects 
(curly lines in Figure~\ref{fig:examplediag}) vanish.

We emphasize the importance of the fact that the gauge 
interactions are decoupled from $\chi_w$ and $h_w$ by 
Wilson lines with the same vector $w^\mu$. 
The production of a resonance in the 
annihilation of two heavy particles is very different in 
this respect from the production in the high-energy 
collision of massless particles, where non-factorizable 
soft effects do exist. The difference is that for such 
a process the decoupling from the Lagrangian leads to 
Wilson lines $Y_{n_+}$, $Y_{n_-}$ and $Y_w$ in different directions, 
where the 
first two are related to the light-like directions 
of the colliding particles, and the production operator 
is not simply a gauge-transformation of itself after 
decoupling.

The factorization of the forward-scattering amplitude 
\eqref{eq:matrix_elem} proceeds as follows. The absence 
of any long-distance  
interaction between the DM and the resonance field 
allows us to split the matrix elements into the two 
factors 
\begin{eqnarray}
&&\int d^4 x \,
\langle \chi \bar \chi|T\{ i J^\dagger(x) iJ(0)\} | \chi \bar \chi \rangle 
\\
&&= i^2 \frac{C^2}{2 m_h} y_{abc}^\dagger y_{a^\prime b^\prime c^\prime} \int d^4 x\,\langle \chi \bar \chi |T\left\{[\chi_w^{\dagger a} \chi_w^b]^\dagger(x) [\chi_w^{\dagger a^\prime} \chi_w^{b^\prime}](0) \right\} | \chi \bar \chi \rangle 
\,\langle 0 |T \left\{ h^{\dagger c}_w(x) h^{c^\prime}_w(0) \right\} | 0 \rangle \,.\nonumber
\label{eq:tproduct}
\end{eqnarray}
The matrix element of non-relativistic fields can now 
be evaluated as in the standard treatment of the 
Sommerfeld effect \cite{Beneke:2014gja}, resulting in 
\begin{eqnarray}
&&\langle \chi \bar \chi |T\left\{[\chi_w^{\dagger a} \chi_w^b]^\dagger(x) [\chi_w^{\dagger a^\prime} \chi_w^{b^\prime}](0) \right\} | \chi \bar \chi \rangle = 
e^{i E t}\,\langle \chi \bar \chi |[\chi_w^{\dagger a} \chi_w^b]^\dagger(0)|0\rangle\,\langle0| [\chi_w^{\dagger a^\prime} \chi_w^{b^\prime}](0) | \chi \bar \chi \rangle
\nonumber\\
&& = 
e^{i E t}\,\big[\psi^{ab}_{E}(0)\big]^* \psi_{E}^{a^\prime b^\prime}(0) 
\times \mbox{Born} 
\equiv e^{i E t}\,S_{\rm SF}(v)\times \mbox{Born}\,,
\end{eqnarray}
neglecting corrections of order $v^2$. In the second line 
we identified the Sommerfeld factor $S_{\rm SF}(v)$, 
which depends on 
$v$ through $E=m_\chi v^2$ in terms of the wave-function 
at the origin of the DM  two-particle scattering 
state in the Yukawa potential. Substituting into 
\eqref{eq:tproduct}, leaves the resonance propagator 
\begin{equation}
 \int d^4 x\,e^{i E t}\,
\langle 0 |T \{ h^{\dagger c}_w(x) h^{c^\prime}_w(0) \} 
| 0 \rangle
=\frac{i\,\delta^{cc^\prime}}{E-\delta M+i\frac{\Gamma_h}{2}}\,.
\end{equation}
Putting everything together and taking the imaginary part, 
the total DM annihilation cross section through the resonance 
can be written as\footnote{The DM particles can also annihilate into 
gauge bosons $XX$. This is formally part of the non-resonant 
cross section, but not necessarily small, since this 
annihilation rate is controlled by the gauge coupling, 
which is independent of the DM coupling $\lambda$ to the resonance. 
See Sec.~\ref{sec:models} below.} 
\begin{equation}
\sigma v_{\rm rel} =  \frac{C^2 Y}{4m_\chi^2} \,S_{\rm SF}(v)\, 
R(v)\,.
\label{eq:sigmaresult}
\end{equation}
Here $Y$ is a schematic notation for the projection of 
$y_{abc}^\dagger y_{a^\prime b^\prime c}$ on the irreducible 
``colour'' channel of the $\chi\chi$ state with respect to 
the gauge symmetry $G$. The value of the Sommerfeld factor is also 
channel-dependent. We further introduced the dimensionless 
quantity\footnote{The total width is 
written as $\Gamma_h =\hat{\Gamma}_h + 
\Gamma_{h\to\chi\chi}$, and in the numerator of 
$R(v)$ only the partial width 
$\hat{\Gamma}_h$  contributes to $\sigma v_{\rm rel}$ 
in the Boltzmann equation.}
\begin{equation}
R(v) = \frac{\frac{m_\chi\hat{\Gamma}_h}{2}}{
(m_\chi v^2-\delta M)^2+
\frac{\Gamma_h^2}{4}}\,,
\end{equation}
which quantifies the resonant enhancement normalized to 
a point-like interaction. 
The annihilation cross section now contains two separate 
enhancements and the factorization of the two long-distance 
effects is manifest in \eqref{eq:sigmaresult}. If 
both enhancements overlap in some velocity region, one 
obtains ``super-resonant'' annihilation cross sections.
A few simple scenarios that can exhibit such behaviour will be 
considered below. 

We briefly discuss the validity of the partial-wave 
expansion, which is equivalent to the expansion in 
the relative velocity of the annihilating DM particles, 
or $E/m_\chi$. The cross section is a function of 
$m_\chi$, $E$ and $\delta M, \Gamma_h$. In the presence of 
a resonance, however, the cross section cannot be 
expanded in $E$ and one may ask whether the expansion of 
the Sommerfeld effect in partial waves is justified. 
Factorization solves this problem as well, since it 
separates the problem into hard-matching coefficients
and the factorized low-energy matrix elements. The computation 
of the Sommerfeld effect is insensitive to the 
existence of the resonance. In the non-relativistic theory, 
the partial-wave expansion can then be constructed by extending 
the matching of the interaction \eqref{eq:interaction} to 
higher orders in $v^2$, in terms of the series 
\begin{equation} 
\lambda \,y_{abc} \,\bar{\chi}_a\chi_b H_c = 
\sum_n \frac{C_ny_{abc}}{m_\chi^{2n}} \,\chi_{wa}^\dagger 
 \big(-\frac{i}{2}
\overleftrightarrow{\bff{\partial}}\big)^{2n}
\chi_{wb} h_{wc}\,.
\end{equation}
In the above discussion we already anticipated that the 
term $n=0$ gives the leading contribution in $E/m_\chi$.

It is well-known that the Sommerfeld 
factor for the Yukawa potential exhibits resonances near 
specific DM mass values, which violate partial-wave 
unitarity due to the presence of a zero-energy bound-state 
\cite{March-Russell:2008klu,Blum:2016nrz}. 
Unitarity is self-consistently restored by including 
the local potential $\delta^{(3)}(\bff{r})$ \cite{Blum:2016nrz} into the 
PNRDM Lagrangian \eqref{eq:PNRDM} which is generated by 
the DM forward $\chi\chi\to\chi\chi$ scattering  
amplitude, whose imaginary part is related to the 
total annihilation cross section. One may wonder 
what are the consequences of \eqref{eq:sigmaresult} 
for unitarity violation. We first note that unitarity 
is not violated by the intermediate-resonance enhancement 
$R(v)$ in the absence of the Sommerfeld factor, 
$S_{\rm SF}(v)=1$. To see this, we remark that the 
approximation of the imaginary part of the $H$ self-energy 
by the on-shell width is not justified when there  
is a decay final-state close to threshold \cite{Duch:2017nbe}, as is the case 
here for $h\to \chi\chi$. In this case, one must use 
$\Gamma(p^2) \propto \sqrt{p^2-4 m_\chi^2}$, where 
$p^2=(2m_\chi+E)^2$ is the off-shell momentum squared 
of the resonance. We can therefore write $\Gamma_h$ 
as 
\begin{equation}
\Gamma_h=\frac{m_\chi}{8\pi}\left(\lambda_{\rm eff}^2 + 
C^2 Y v_{\rm rel} \right) \equiv \hat{\Gamma}_h + 
\Gamma_{h\to\chi\chi},
\end{equation}
where $\lambda_{\rm eff}^2$ parameterizes the coupling to 
all final states in $h$ decay, which are not close to 
threshold. Then from \eqref{eq:sigmaresult}
\begin{eqnarray}
\sigma v_{\rm rel}\big|_{\mbox{\rm \tiny no SF}} 
\leq  \frac{C^2 Y}{4m_\chi^2}\,\frac{2m_\chi\hat{\Gamma}_h}
{\Gamma_h^2}
= \frac{4\pi}{m_\chi^2 v_{\rm rel}} \,
\frac{\lambda_{\rm eff}^2\times C^2 Y v_{\rm rel}}{(\lambda_{\rm eff}^2 + 
C^2 Y v_{\rm rel})^2} \leq  
\frac{\pi}{m_\chi^2 v_{\rm rel}}\,,
\end{eqnarray}
which remains a factor of 4 below the unitarity bound, 
hence (S-wave) partial unitarity is guaranteed.\footnote{
The factor of four arises from the 
initial-state spin-average and the fact that only one of 
the four possible initial spin states couples to the scalar 
resonance.}

Nevertheless, when $R(v)\gg 1$ the unitarity violation 
from $S_{\rm SF}(v)$ near certain resonant DM mass 
values is enhanced and operative in a wider DM mass window 
around the resonant value. Yet, the basic mechanism of unitarity restoration remains 
the same as in the standard case, because the intermediate 
resonance that causes the enhancement $R(v)$ also 
contributes to the imaginary part of the $\chi\chi$ 
forward scattering amplitude. In contrast to the 
standard situation, however, the forward-scattering 
amplitude does not match to a local potential in 
the non-relativistic effective theory, since resonant 
scattering is a long-distance process. Formally integrating 
out the resonance, results in the temporally 
non-local operator
\begin{equation}
- \mbox{const.}\times 
\int_0^\infty ds \,e^{-i (\delta M-i \Gamma_h/2) s}\,
\left[\chi_w^\dagger \chi_w\right](t+s, \mathbf{x}) \left[\chi_w^\dagger \chi_w\right](t, \mathbf{x})\,.
\label{eq:nonlocalpotential}
\end{equation}
The PNRDM field equation then no longer takes the 
form of a Schr\"odinger equation, and the solution for 
the unitarized annihilation cross section becomes  
complicated.

\begin{figure}[t]
\centering
\includegraphics[width=0.47\textwidth]{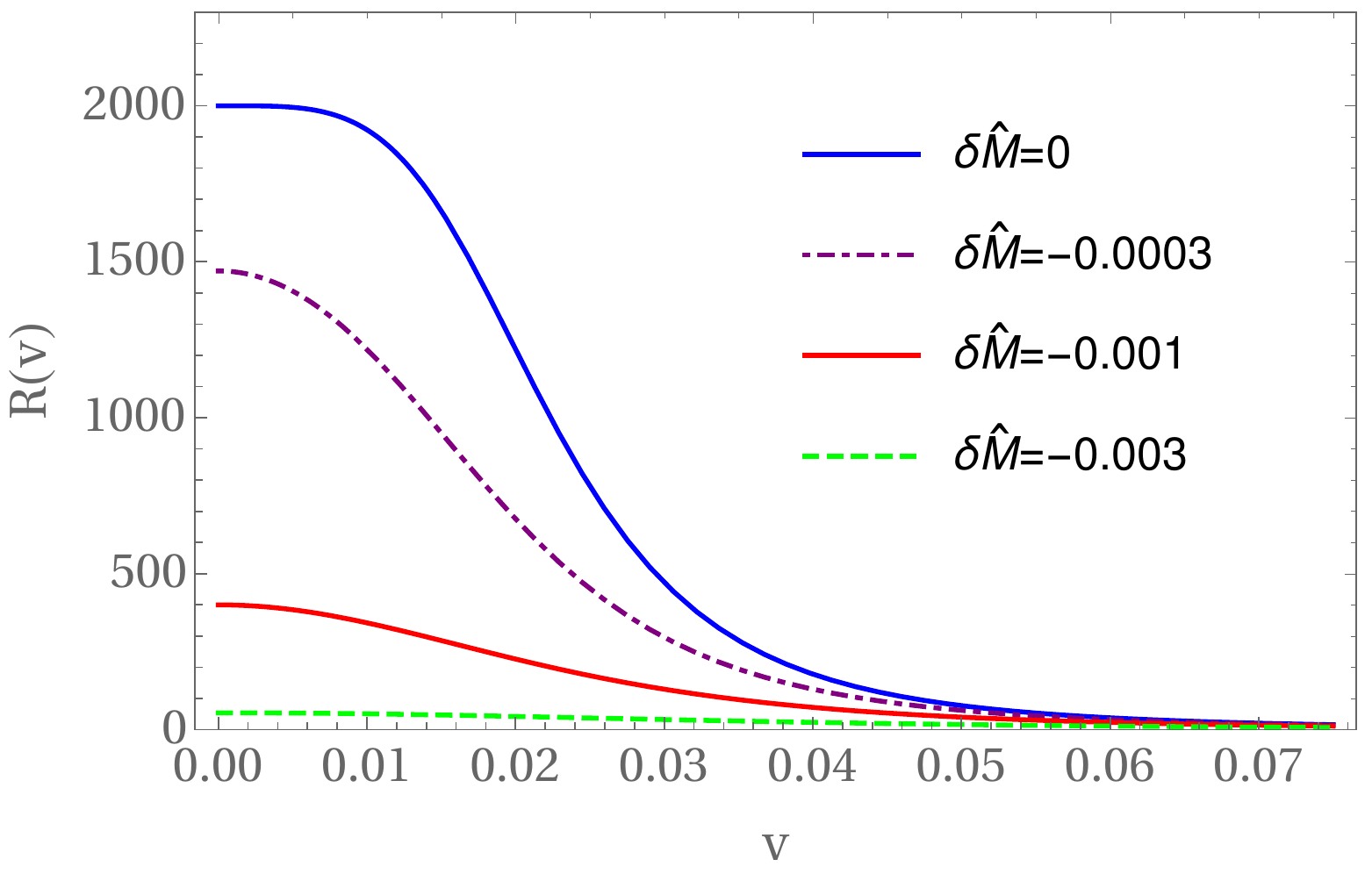}
\hskip0.2cm\includegraphics[width=0.47\textwidth]{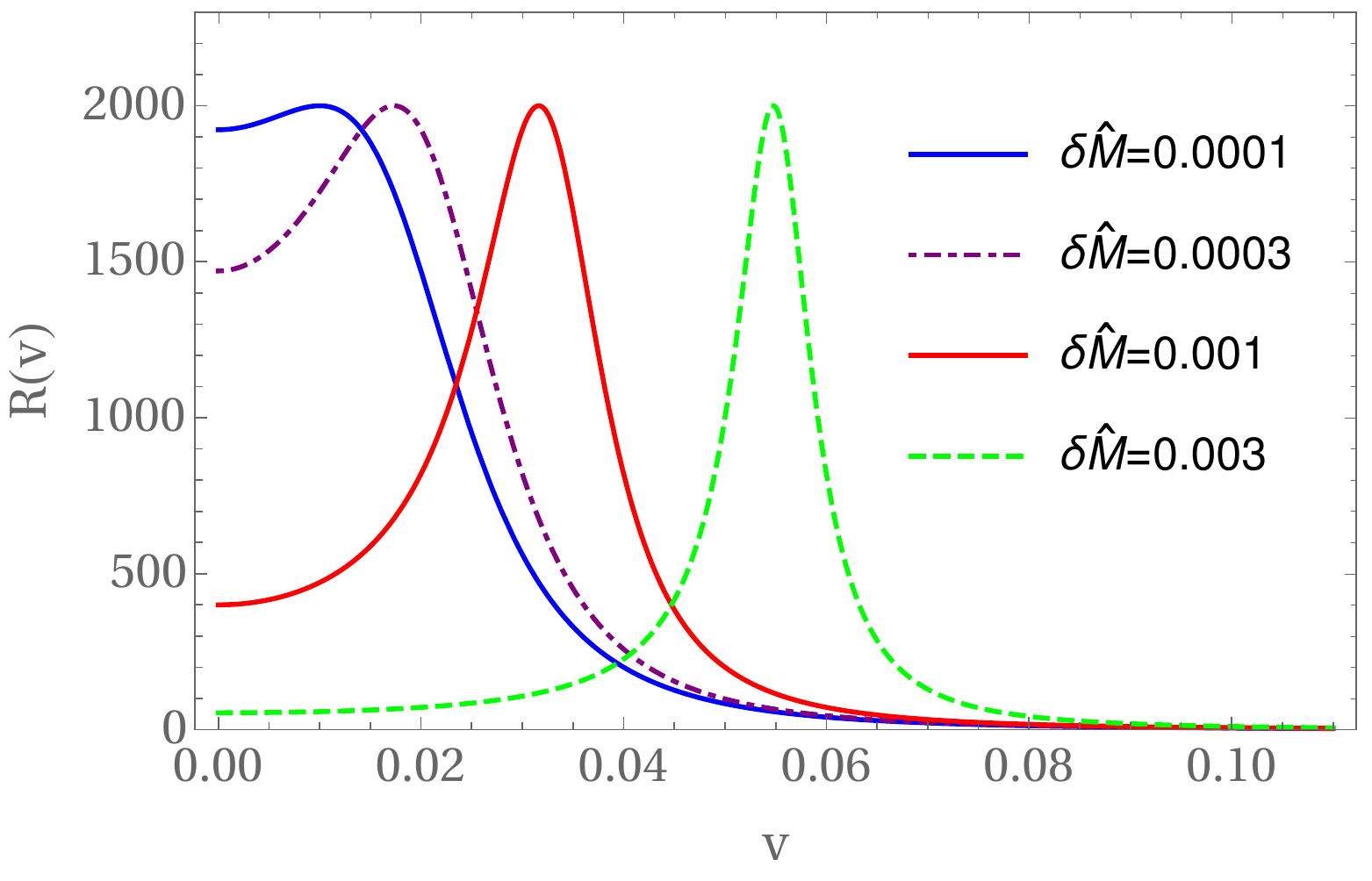}
\caption{\small Velocity dependence of the resonant 
enhancement $R(v)$ for $\delta M\leq 0$ (left) and 
$\delta M>0$ (right) for $\Gamma_h/m_\chi=0.001$ and various 
values of $\delta\hat{M}=\delta M/m_\chi$.}
\label{fig:rv}
\end{figure}

It is useful to gain some qualitative understanding of the 
velocity dependence of $R(v)$, given $\delta M$ and 
$\Gamma_h/m_h \approx \Gamma_h/(2 m_\chi)$.\footnote{
For simplicity, we set $\Gamma_h$ equal to 
 $\hat{\Gamma}_h$ for this discussion.} We first note 
that for $\delta M<0$ the resonance is lighter than 
$2 m_\chi$. In this case, the annihilation occurs 
always beyond the resonance peak of the propagator 
and the resonance enhancement is monotonically decreasing 
as $v$ grows (Figure~\ref{fig:rv}, left). When the velocity 
decreases, $R$ first increases as $1/v^4$ with maximal 
slope for $v^2\approx 0.88 \,\Gamma_h/(2m_\chi)$ (as long as 
$|\delta M|\,\lessim\, \Gamma_h$), and then saturates at 
the maximal value 
\begin{equation}
R(0) = \frac{\frac{m_\chi\Gamma_h}{2}}{
\delta M^2+\frac{\Gamma_h^2}{4}} < R_{\rm max} = 
\frac{2m_\chi}{\Gamma_h}\,.
\end{equation} 
Interestingly, this behaviour is similar to the saturation 
of the Sommerfeld effect in the Yukawa potential. In the 
latter case the saturation value depends on the strength 
of the potential and the value of $m_\chi/m_X$.

For positive $\delta M$, the DM particles must have 
finite velocity $v_{\rm peak}^2=\delta M/m_\chi$ to produce 
the resonance exactly on its peak. At this velocity, 
the maximally possible enhancement 
$ R_{\rm max}$ is attained independent 
of the value of $\delta M$  (Figure~\ref{fig:rv}, right). For smaller velocity, 
$R(v)$ drops to the same $R(0)$ as for the corresponding 
negative $\delta M$. 
The annihilation cross section can now be 
dramatically increased in a narrow velocity interval, 
which is not possible for Sommerfeld enhancement. 
However, when $\delta M \ll \Gamma_h$, the peak becomes increasingly 
one-sided, since it is regularized by the finite width. 
A pronounced resonance peak can appear in the velocity spectrum 
only for 
$v_{\rm peak}^2 \,\grtsim\, \Gamma_h/(2 m_\chi)$.
 
The velocity peak can boost 
freeze-out relative to late time annihilation, which occurs 
at smaller $v$, or vice versa \cite{Ibe:2008ye}, 
or bias indirect DM detection from astrophysical objects with markedly 
different velocity distribution, if the resonance is 
sufficiently narrow. 


\section{Resonant dark matter scenarios}
\label{sec:models}

DM annihilation through a resonance is by itself a 
long-known mechanism to boost the annihilation rate 
\cite{Griest:1990kh,Gondolo:1990dk,Jungman:1995df}, 
but to our knowledge the coaction of resonant and 
Sommerfeld effect has not been considered up to 
now, even leaving aside radiative effects. In the 
following, we discuss two potentially interesting 
scenarios, focusing on the question whether super-resonant 
behaviour is possible.

\subsection{Higgs portal with dark sector mediator}
\label{sec:Higgsportal}

We first study the SM Higgs portal model with a complex scalar 
DM particle~$S$ of mass $m_S$, augmented by a dark-sector U(1)$_X$ 
symmetry, which is spontaneously broken to give the dark gauge 
boson a small mass $m_X\ll m_S$. The relevant 
terms in the Lagrangian are 
\begin{align}
\mathcal{L} &\supset
 \left(D_\mu S\right)^\dagger \left(D^\mu S\right) - m_S^2 |S|^2 - \frac{1}{4} X_{\mu \nu} X^{\mu \nu}  -  \frac{\lambda_S}{4} |S|^4 + \lambda_{HS} (H^\dagger H) |S|^2 + \mathcal{L}_{\rm mix} \label{eq:DM_model}
\end{align}
where $H$ is the SM Higgs doublet. The mass and width of the 
Higgs particle $h$ are given by $m_h=125\,$GeV and 
$\Gamma_{h,\rm SM}=4.1\,$MeV, 
respectively, and since near resonance the mass of the DM 
scalar satisfies $m_S \approx m_h/2$, the only adjustable 
parameter related to the resonance is 
$\delta M= m_h-2 m_S$. 

To be cosmologically viable, the light dark-sector gauge boson 
must decay into SM model particles. After electroweak symmetry 
breaking, $X$ mixes with the SM gauge bosons through the terms 
\begin{align}
\mathcal{L}_{\rm mix} = \frac{\epsilon_\gamma}{2} X_{\mu \nu} F^{\mu \nu} + \epsilon_Z m_Z^2 X_{\mu} Z^\mu
\end{align}
with $\epsilon_\gamma, \epsilon_Z \ll 1$. The resulting decay 
width of $X$ into photons is given by \cite{Kaplinghat:2013yxa}
\begin{align}
\Gamma_{X \to \gamma} = \alpha_{\rm em} m_X \epsilon_\gamma^2/3 \, .
\end{align}
Choosing $\epsilon_{\gamma},\epsilon_Z = 10^{-9}$ ensures dark 
gauge boson decay before nucleosynthesis. Additionally, 
constraints from supernova cooling require $m_X \,\grtsim \, 
100\,{\rm MeV}$ \cite{Dent:2012mx}. 

The dark gauge boson is still abundant during freeze-out, and 
the process $SS^* \leftrightarrow XX$ must be considered. 
For $m_S \gg m_X$, the annihilation rate is
\begin{align}
\sigma v_{\rm rel}(SS^* \to XX) = \frac{4\pi \alpha_X^2}{m_S^2}
\times S_{\rm SF}(v)\,,
\label{eq:SStoXX}
\end{align}
where $S_{\rm SF}(v)$ is the Sommerfeld factor generated by 
the dark U(1) force. 
Since $m_S\approx 62.5\,$GeV is fixed, this implies that the 
dark-sector gauge coupling $g_X=\sqrt{4\pi\alpha_X}$ cannot be 
too large, since then the relic density would be 
under-abundant compared to the observed $\Omega h^2=0.120$.

The model is subject to the same constraints as the ordinary Higgs 
portal with a real singlet scalar (and no dark mediator), which 
we do not repeat here. 
Suffice it to say that the resonant regime is in the best-fit 
region \cite{GAMBIT:2017gge}. It remains to show, however, that 
additional processes induced by the dark gauge boson do not lead to further constraints. DM  
loops generate an effective interaction with the 
SM Higgs boson, given by 
\begin{equation}
\mathcal{L}_{\rm EFT} \supset 
\sqrt{2}\lambda_{XXH}\,\vev \,h X_{\mu \nu} X^{\mu \nu} 
\quad\mbox{with}\quad
\lambda_{XXH} = -\frac{\alpha_X \lambda_{HS}}{48 \pi m_S^2} \,,
\end{equation}  
where $\vev \approx 246\,$GeV denotes the SM Higgs vacuum expectation 
value. This results in a contribution to the invisible Higgs 
width, 
\begin{equation}
\Gamma(h \to XX) = 
\frac{\lambda_{HS}^2 \alpha_X^2 m_h^3 \vev^2}{4608 \pi^3 m_S^4} \,.
\end{equation}  
If $m_h>2 m_S$, the channel 
\begin{equation}
\Gamma(h \to SS^*) = \frac{\lambda_{HS}^2 \vev^2 }{8 \pi m_h} 
\sqrt{1 -\frac{4 m_S^2}{m_h^2}}
\end{equation}   
contributes directly to the Higgs width, but is phase-space 
suppressed near resonance. However, this decay now receives a 
final-state Sommerfeld enhancement from $X$-exchange. We 
checked that for scenarios of potential interest, the values of 
the coupling $\lambda_{HS}$ are such that the invisible 
Higgs width remains far below the SM width.

The annihilation cross sections relevant for setting the relic 
abundance of $S$ and for indirect detection signals are \eqref{eq:SStoXX} and the resonant
 annihilation of two DM particles via the SM Higgs boson
\begin{align}
\sigma v_{\rm rel} (SS^* \to h \to {\rm SM \, SM}) = \frac{\lambda_{HS}^2 \vev^2}{4 m_S^3}  \frac{\frac{1}{2} \Gamma_{h,{\rm SM}}}{\left(E - \delta M\right)^2 + \frac{\Gamma_{h}^2}{4}}\times S_{\rm SF}(v)\,.
\end{align}
Here 
$E=m_S v^2$ is the kinetic energy of the two-particle $SS^*$ state and $\Gamma_h = \Gamma_{h, {\rm SM}} + \Gamma_{h,{\rm inv.}}$ the total Higgs decay width including the invisible one. The 
resonance enhancement is determined by $\Gamma_h$ and 
$\delta M$, while Sommerfeld enhancement requires 
\begin{equation}
\frac{\pi \alpha_X m_S}{m_X} \geq 1 \,.
\end{equation}
In the present model, the combination of parameters on the left-hand side cannot be made arbitrarily large, since i) $\alpha_X$ is 
constrained to values below $10^{-3}$ by the correct relic abundance 
requirement, which limits $\sigma v_{\rm rel}(SS^* \to XX)$ 
given in \eqref{eq:SStoXX}, and ii)  $m_X \,\grtsim \, 
100\,{\rm MeV}$. Thus $\pi \alpha_X m_S/m_X\sim 1$ can 
be achieved, if $m_X$ and $\alpha_X$ 
are pushed towards their limits, but not much larger values. However, the first Sommerfeld 
resonance in the Yukawa potential which appears at 
$\alpha_X\approx  \frac{\pi^2}{6}\times m_X/m_S \;\grtsim\; 
2.5\cdot 10^{-3}$, cannot be reached. Hence, only a moderate 
$\mathcal{O}(1)$ Sommerfeld enhancement is to be expected, and 
it is therefore not possible to obtain 
truly super-resonant enhancements in the SM Higgs-portal model with a 
dark-sector gauge boson. 

\begin{figure}[t]
\centering
\includegraphics[width=0.65\textwidth]{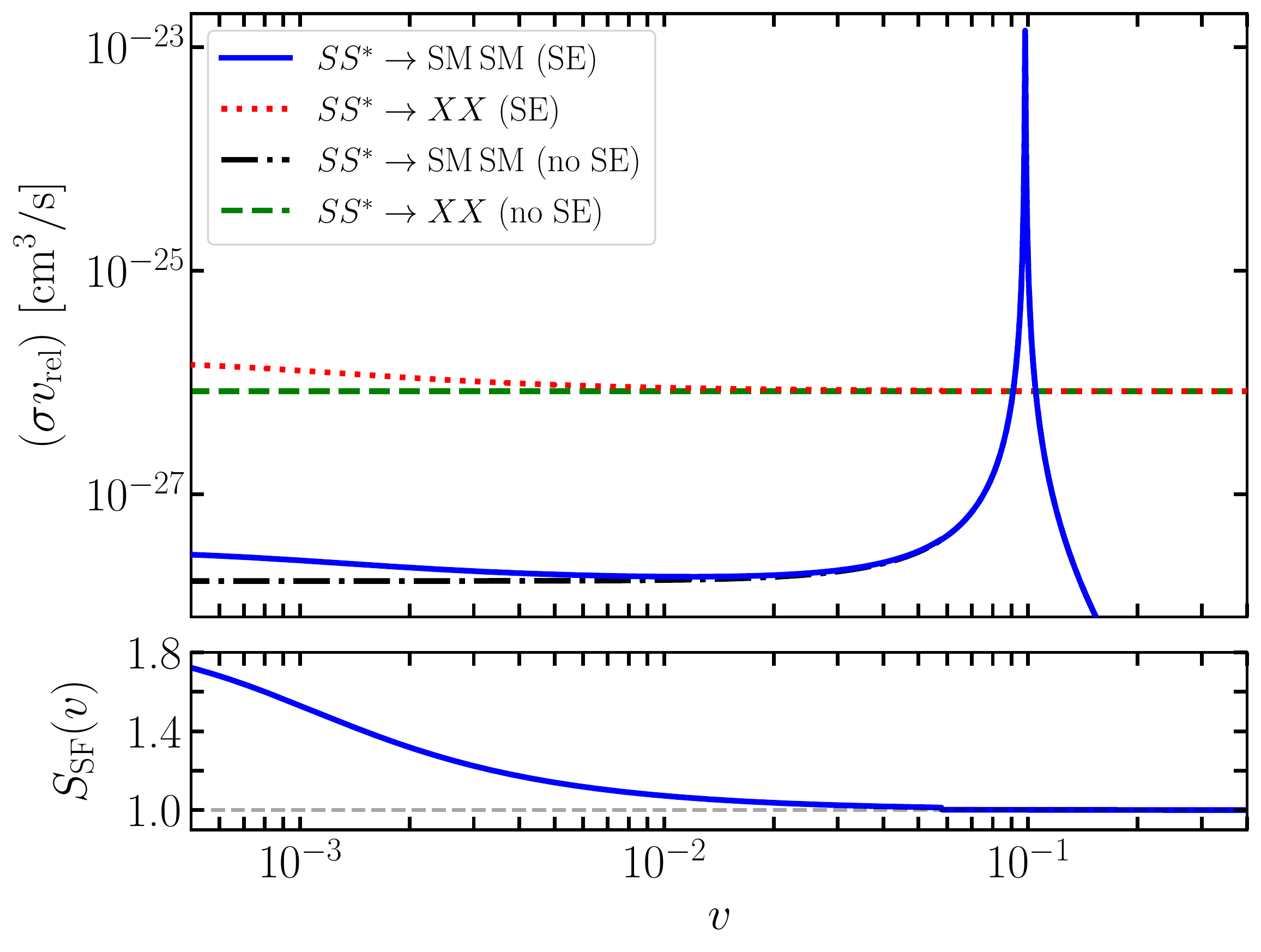}
\includegraphics[width=0.67\textwidth]{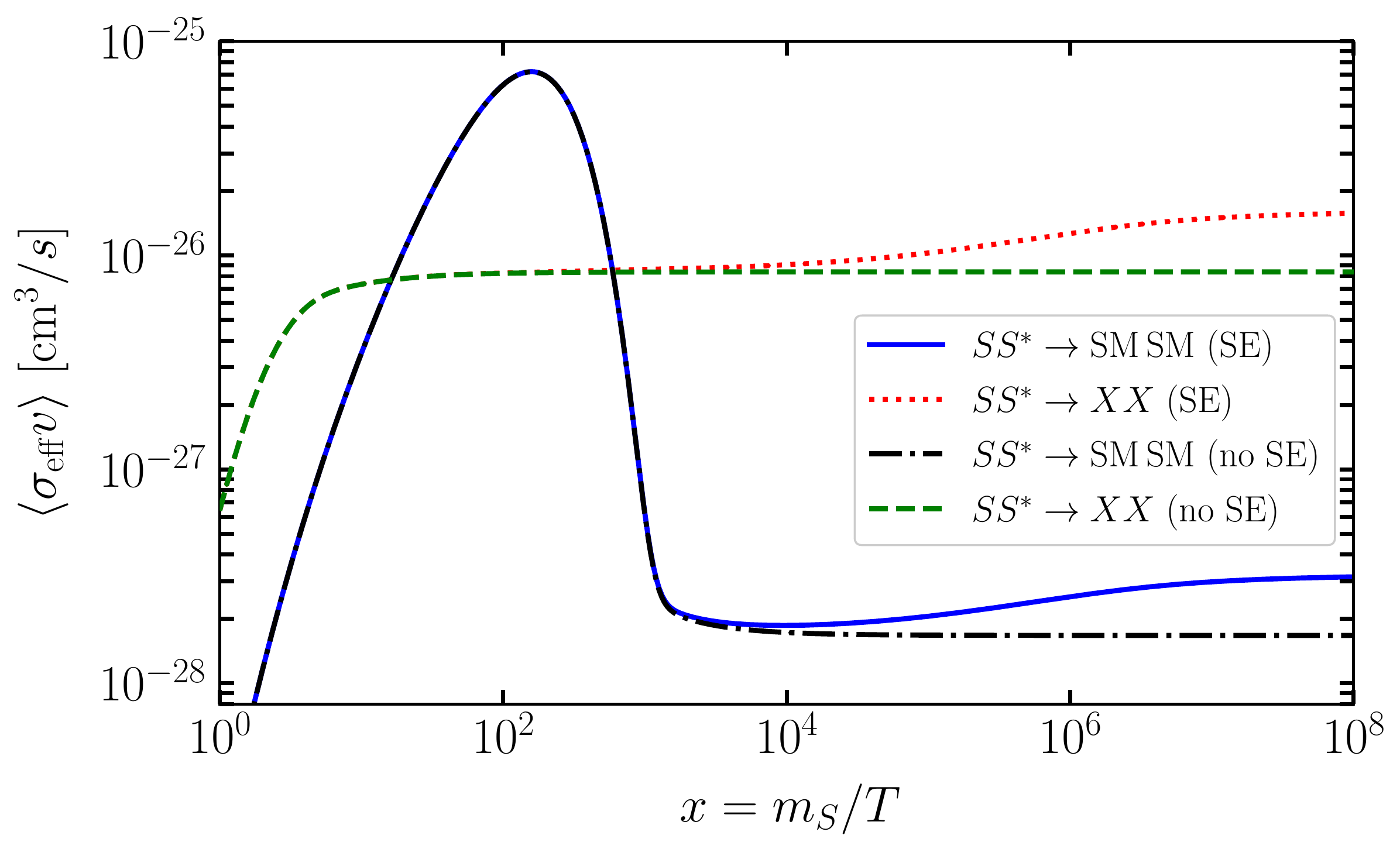}
\caption{Upper panel: Velocity dependence of the relevant DM annihilation processes with and without Sommerfeld 
enhancement (ratio subtended) in the SM Higgs-portal scenario. 
Lower panel: Thermally averaged annihilation cross section.}
\label{fig:higgsportal_benchmark}
\end{figure}

An example of a viable scenario that accounts for 
$\Omega_{\rm DM} h^2=0.120$ uses $m_S=62.2\,$GeV, corresponding to 
$\delta M=0.6\,$GeV, $m_X=0.1\,$GeV, $\alpha_X=4.7\cdot 10^{-4}$ 
and the DM coupling to the Higgs resonance $\lambda_{HS} = 
2\cdot 10^{-4}$. For these parameter values the spin-independent direct detection cross section is \cite{GAMBIT:2017gge}
\begin{equation}
\sigma_{\rm SI}=2\,\frac{m_N^4}{4\pi(m_S+m_N)^2}
\frac{(2\lambda_{HS})^2 f_N^2}{m_h^4}\approx 8\cdot 
10^{-49}\,\mbox{cm}^2\,, 
\end{equation}
about two orders of magnitude below the present limits 
\cite{LZ:2022ufs}. The value of $\delta M>0$ is chosen 
such that a narrow resonant enhancement appears for velocity 
$v\approx 0.1$, which boosts the annihilation into SM 
particles when freeze-out happens (upper panel, 
Figure~\ref{fig:higgsportal_benchmark}). Except for this 
narrow region, the annihilation cross section is dominated 
by the annihilation into a pair of dark gauge bosons. The 
impact on the thermally averaged cross section is clearly 
visible in the lower panel of the figure. The annihilation 
rate at small velocity relevant today is small and evades 
the constraints, which mainly arise from the Fermi-LAT, 
MAGIC \cite{MAGIC:2016xys} and  
H.E.S.S. \cite{HESS:2022ygk} experiments' 
non-observation of annihilation into the $b\bar{b}$ 
final state. The Sommerfeld enhancement reaches nearly a 
factor of two (subtended panel), but, as expected, does not 
cause a dramatic increase of cosmic-ray signals. 

As a general remark, we note that scenarios with $\delta M<0$ 
or small positive $\delta M\,\lessim\,\Gamma_h$ that satisfy 
$\Omega h^2=0.120$ are likely to 
produce too large annihilation cross sections at small $v$ 
and are therefore strongly constrained by the absence of indirect 
detection signals.

\subsection{MSSM template}

The large MSSM parameter space features both the Sommerfeld 
enhancement through the standard electroweak gauge forces and 
resonant annihilation. The mass of the $W$ boson and the 
size of the SU(2) and U(1)$_Y$ gauge couplings requires DM masses 
of a few TeV or larger in order to have an $\mathcal{O}(1)$ 
Sommerfeld effect. Resonant annihilation then requires a decoupling 
scenario with a SM-like Higgs boson and a heavy Higgs doublet 
consisting of nearly degenerate $A^0$, $H^0$ and $H^\pm$ 
bosons. The dominant decay channels of the heavy Higgs bosons are 
into gauge bosons, top quarks, and bottom quarks, depending on 
the value of $\mbox{tan}\,\beta$. The $A^0$ and $H^0$ resonances 
may overlap, when the mass splitting is small enough. In the 
following simplified description, we neglect the interference 
effects that can occur in this situation and assume a single 
resonant intermediate state.

The treatment of the Sommerfeld effect in the full MSSM can 
be found in \cite{Beneke:2012tg,Beneke:2014gja}. In these works 
the case of resonant annhilation was excluded, since the 
Born cross sections were expanded in $E$. With the result of the 
previous section, this restriction 
could now be removed. Leaving a discussion of the full MSSM 
to future work, we consider here a grossly simplified template 
for resonant neutralino annhilation in the MSSM, consisting 
of the neutralino $\chi$, a heavy Higgs boson $A^0$ (not necessarily 
CP-odd) with mass close to $2 m_\chi$ and a massive U(1) gauge 
boson $X$ with Lagrangian
\begin{align}
\mathcal{L} = \overline{\chi} (i \slashed{D} - m_\chi) \chi - \frac{1}{4} X_{\mu \nu} X^{\mu \nu} + y A^0 \overline{\chi} \chi - \frac{1}{2} m_A^2 \left(A^0\right)^2+ \mathcal{L}_{\rm SM}\,.
\end{align}
To allow for a $U(1)$ force the ``neutralino'' is assumed to be a 
Dirac fermion. 
The relevant DM annihilation cross sections are 
\begin{eqnarray}
\sigma v_{\rm rel} (\chi \bar \chi \to A^0 \to  {\rm SM \, SM}) &=& \frac{y^2}{2 m_\chi} \frac{\frac{1}{2} \Gamma_{A, {\rm SM}}}{(m_\chi v^2 - \delta M)^2 + \frac{\Gamma_{A}^2}{4}}\,, 
\label{eq:MSSMresonantrate}\\
\sigma v_{\rm rel} (\chi \bar \chi \to XX) &=& \frac{\pi \alpha_X^2}{m_\chi^2} \,.
\label{eq:MSSMgaugebosonrate}
\end{eqnarray}
For multi-TeV Higgs bosons in the MSSM, their decay width into 
SM particles typically results in $\Gamma_{A, {\rm SM}}/m_A 
\sim 10^{-3} \ldots 10^{-2}$. Since the neutralino-Higgs coupling 
$y$ is related to the electroweak gauge couplings, the decay 
rate for $A^0\to\chi\bar \chi$,  
\begin{equation}
\Gamma(A^0 \to \chi \bar \chi) = \frac{y^2 m_\chi^2}{2 \pi m_A} \sqrt{1- \frac{4 m_\chi^2}{m_A^2}}\,\theta(m_A - 2 m_\chi)\,,
\end{equation}
can be neglected due to its phase-space suppression
even in the presence of the final-state Sommerfeld effect, except 
near a Sommerfeld-resonant $m_\chi$ mass value. 
Nevertheless, the numerical evaluations below include this 
decay channel. 

The Sommerfeld effect is important in the MSSM for a variety 
of situations (see, for example, \cite{Beneke:2014hja,Beneke:2016ync,Beneke:2016jpw,Hryczuk:2019nql}). We are interested here in 
models that yield the observed relic abundance through a standard 
thermal freeze-out. In this case, Sommerfeld enhancement is 
particularly important for wino-dominated neutralinos, 
which can have mass between about 2 and 3.5~TeV 
\cite{Beneke:2016ync}. Much (but not all) of this parameter space is, 
however, already excluded by the absence of indirect detection 
signals \cite{Beneke:2016jpw,Hryczuk:2019nql} even when assuming 
a cored DM profile of the Milky Way. Adding a resonant 
component \eqref{eq:MSSMresonantrate} to the already large 
annihilation rate into gauge bosons \eqref{eq:MSSMgaugebosonrate} that enhances late-time annihilation ($\delta M<0$ or 
$\delta M\lessim \Gamma_h$) worsens this tension. 
Bino-dominated DM, on the 
other hand, does not exhibit a significant Sommerfeld effect.

An interesting situation may arise for the wino- or 
Higgsino-dominated neutralino when $\delta M > \Gamma_h$. 
We consider explicitly a Higgsino-like setting in the template model in the following. In the pure Higgsino (minimal SU(2) doublet) model, the correct relic abundance is obtained for $m_\chi \approx 
1.1~$TeV, at which mass the Sommerfeld enhancement is only a few percent effect on the relic density and about a factor of two for small velocities $v\approx 10^{-3}$ \cite{Hisano:2004ds,Beneke:2019gtg}. This allows for the possibility to raise the neutralino 
mass while avoiding over-abundance by resonant annihilation 
through a resonance peak in the velocity spectrum at 
early times. 
The annihilation cross section in the present universe then 
experiences super-resonant enhancement by a sizable Sommerfeld 
effect (due to the large mass) and off-peak resonant 
enhancement. In the full MSSM this requires a bino admixture 
to the dominantly Higgsino-neutralino to generate the DM coupling 
to the heavy Higgs boson. 

\begin{figure}[t]
\centering
\includegraphics[width=0.65\textwidth]{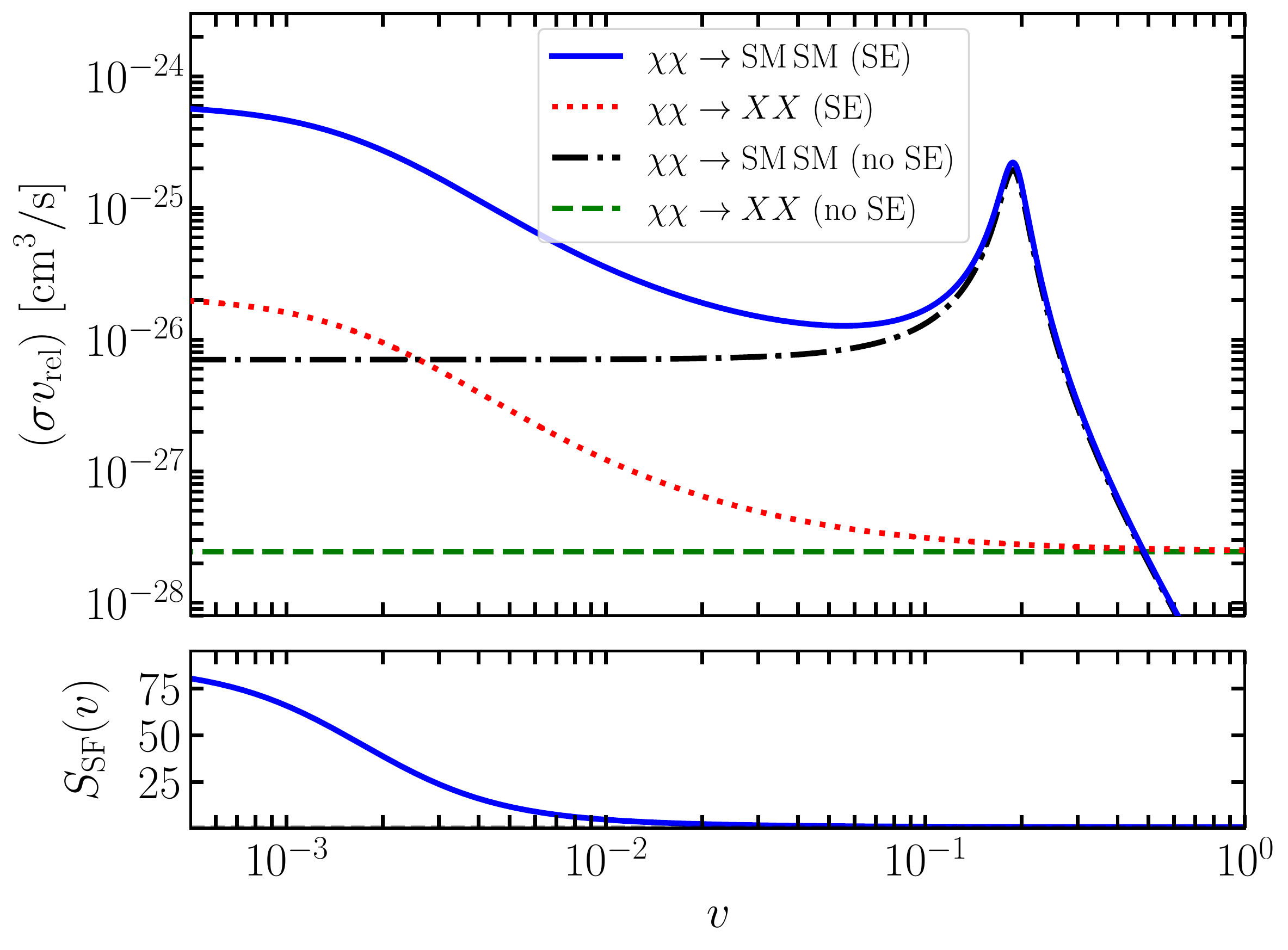}
\includegraphics[width=0.67\textwidth]{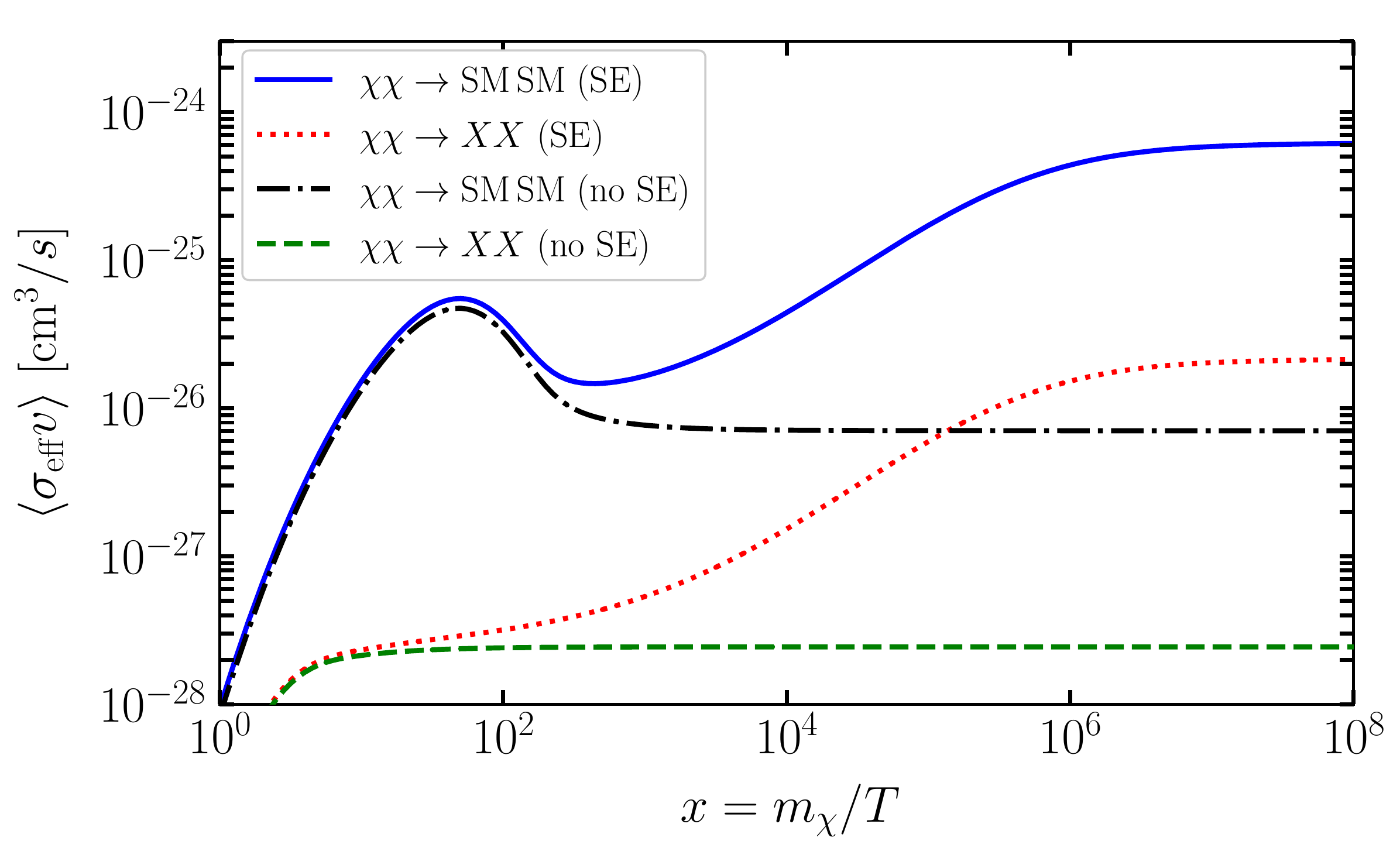}
\caption{Upper panel: Velocity dependence of the relevant DM annihilation processes with and without Sommerfeld 
enhancement (ratio subtended)  in the MSSM template scenario. 
Lower panel: Thermally averaged annihilation cross section.}
\label{fig:mssm_benchmark}
\end{figure}

We mimic this situation in our 
template model by choosing the effective gauge coupling 
$\alpha_X = \frac{1}{2} \alpha_2(m_Z)$ and $y = 
\sqrt{4 \pi  \alpha_2(m_Z)} \cdot 0.152$, where 
$\alpha_2(m_Z) = 1/29.792$ is the standard SU(2) gauge 
coupling. We further adopt the heavy Higgs width 
$\Gamma_{A, {\rm SM}} =  m_A/150$ and $m_X=m_W=80.385 \,{\rm GeV}$. 
The correct relic abundance $\Omega_{\rm DM} h^2 = 0.120$ is 
now obtained for the significantly larger ``dominantly-Higgsino" 
DM mass 
$m_\chi = 6500 \,{\rm GeV}$ and Higgs mass 
$m_A =13230 \, {\rm GeV}$. In this model $\delta M=230$~GeV 
is positive and exceeds $\Gamma_{A, {\rm SM}} = 88.2~$GeV, 
which leads to a pronounced resonance peak in the 
annihilation cross section at velocity $v\approx 0.2$, 
as shown in Figure~\ref{fig:mssm_benchmark}. The resonant 
enhancement is crucial to sufficiently deplete the DM abundance 
during freeze-out. The Sommerfeld enhancement, on the other hand, is relatively small  
during this period and does not influence the freeze-out 
noticably. However, it reaches $\mathcal{O}(10^2)$ below $v\approx 
10^{-3}$ and boosts the annihilation cross section in the late 
universe to a level that might be observable with future 
cosmic-ray experiments even for a cored DM profile in the 
Galactic Center. While the freeze-out is entirely determined 
by the DM annihilation through the $A^0$ resonance, the 
annihilation into gauge bosons dominates at small velocities 
and late times. 

A similar scenario can be found for wino-like DM in the few TeV mass range between the first and second Sommerfeld resonance.


\section{Conclusion}
\label{sec:conclusion}

In this letter we investigated the factorization properties 
of the DM annihilation cross section when the two long-distance 
effects of resonant annihilation and Sommerfeld enhancement 
operate together. Our main theoretical result is that unlike 
in the production of a resonance in high-energy scattering of 
ultra-relativistic particles, 
non-factorizable long-distance effects cancel completely in 
heavy-particle annihilation up to corrections of higher-order 
in the non-relativistic and small-width expansion. As a consequence, 
the annihilation cross section is the product of hard amplitudes, 
the Sommerfeld factor and the Breit-Wigner factor.

When the resonance is slightly lighter than $2 m_\chi$, or 
$\delta M$ is small compared to the resonance width, the 
Sommerfeld and resonant enhancement both show saturation 
behaviour as $v\to 0$. In such cases, it is difficult to obtain 
super-resonant behaviour from the coaction of both mechanisms, 
since satisfying indirect detection constraints usually leads to 
an overabundance of DM when produced through thermal 
freeze-out.

We specifically analyzed the SM Higgs-portal scalar DM model 
with a dark-sector gauge symmetry to illustrate this point. 
The requirement of resonance fixes the DM mass to $m_h/2$.  
In contrast, the mass of the DM is unconstrained in the MSSM, since the heavy Higgs 
mass is not fixed. Simplifying 
to a template model that features electroweak-size gauge 
couplings and a Yukawa coupling of the neutralino to 
a heavy MSSM Higgs boson, we find the interesting possibility 
of heavy Higgsino- or wino-like DM, whose relic abundance is set 
by resonant annihilation, while the late-time annihilation 
leads to signals, potentially observable in the near future, 
via a strong Sommerfeld effect. This motivates a closer 
inspection of the relevant parameter space in the full MSSM 
along the lines 
of \cite{Beneke:2016ync,Beneke:2016jpw,Hryczuk:2019nql}.  

\vspace*{0.6cm}\noindent
{\bf Acknowledgements.} MB thanks the Albert Einstein 
Center at the University of Bern for hospitality while 
this work was finalized. This research has been supported 
by the DFG Collaborative Research Center ``Neutrinos and Dark
 Matter in Astro- and Particle Physics'' (SFB 1258). 

\providecommand{\href}[2]{#2}\begingroup\raggedright\endgroup

\end{document}